\documentclass[12pt]{article}

\usepackage{epsfig}

\catcode`@=11

\def\@citex[#1]#2{\if@filesw\immediate\write\@auxout{\string\citation{#2}}\fi
  \def\@citea{}\@cite{\@for\@citeb:=#2\do
    {\@citea\def\@citea{,\penalty\@m}\@ifundefined
      {b@\@citeb}{{\bf ?}\@warning
       {Citation `\@citeb' on page \thepage \space undefined}}%
\hbox{\csname b@\@citeb\endcsname}}}{#1}}

\def\citer{\@ifnextchar [{\@tempswatrue\@citexr}{\@tempswafalse\@citexr[]}}

\def\@citexr[#1]#2{\if@filesw\immediate\write\@auxout{\string\citation{#2}}\fi
  \def\@citea{}\@cite{\@for\@citeb:=#2\do
    {\@citea\def\@citea{--\penalty\@m}\@ifundefined
       {b@\@citeb}{{\bf ?}\@warning
       {Citation `\@citeb' on page \thepage \space undefined}}%
\hbox{\csname b@\@citeb\endcsname}}}{#1}}
\catcode`@=12

\def\Journal#1#2#3#4{{#1} {\bf #2}, #3 (#4)}

\def\NPB{{\em Nucl. Phys.} B}
\def\NPPS{\em Nucl. Phys. (Proc. Suppl.)} 
\def\PLB{{\em Phys. Lett.}  B}
\def\PRL{\em Phys. Rev. Lett.}
\def\PRD{{\em Phys. Rev.} D}
\def\ZPC{{\em Z. Phys.} C}

\def\be{\begin{equation}}
\def\ee{\end{equation}}
\def\bea{\begin{eqnarray}}
\def\eea{\end{eqnarray}}

\newcommand{\lwig}{\mbox{\,\raisebox{.3ex}
    {$<$}$\!\!\!\!\!$\raisebox{-.9ex}{$\sim$}\,}}
\newcommand{\gwig}{\mbox{\,\raisebox{.3ex}
    {$>$}$\!\!\!\!\!$\raisebox{-.9ex}{$\sim$}}\,}

\newcommand{\ai}{{\overline{I}}}
\newcommand{\iai}{I\overline{I}} 
\newcommand{\xpr}{{x^\prime}}

\date{}

\begin{document}

\title{{\normalsize\rightline{DESY 98-115}\rightline{hep-ph/9808482}} 
  \vskip 1cm 
  \bf Instanton Searches at HERA\thanks{Talk presented at Quarks `98,  
    10th International Seminar on High Energy Physics,
    Suzdal, Russia, May 1998; to be published in the Proceedings.}
}
\author{A. Ringwald and F. Schrempp\\[0.3cm]
Deutsches Elektronen-Synchrotron DESY, Hamburg, Germany}

\begin{titlepage} 
\maketitle

\vspace{2cm}
\begin{abstract}
The present status of our ongoing systematic study of the discovery potential 
of QCD-instanton induced events in deep-inelastic scattering at HERA is 
briefly reviewed. We emphasize our recent progress in predicting the 
cross-sections of instanton-induced processes. Our finalized
predictions include a dramatic 
improvement of the residual renormalization-scale dependencies and the
specification of a ``fiducial'' kinematical region in the relevant Bjorken 
variables extracted from recent lattice simulations. 
Published upper limits on instanton-induced cross-sections based on 
single observables in the final state such as the flow of strange particles
and the multiplicity distribution of charged particles are already of the
order of our estimate. Thus, a decisive search for instanton-induced
events in deep-inelastic scattering at HERA, based on a multi-observable 
analysis, seems feasible.   
\end{abstract}
\thispagestyle{empty}
\end{titlepage}
\newpage \setcounter{page}{2}

\section{Introduction}\label{s0}

In this contribution, we briefly review the present status of our
ongoing systematic study~\citer{rs-pl,crs} of the discovery potential
of QCD-instanton induced events in deep-inelastic scattering (DIS) at HERA.  

Hard scattering processes in elementary particle physics are successfully 
described by the Standard Model of strong (QCD) and electro-weak (QFD) 
interactions in its perturbative formulation.
However, perturbation theory does not exhaust all
possible hard scattering processes.  

Instantons~\cite{bpst}, 
fluctuations of (non-abelian) gau\-ge 
fields representing topology changing tunnelling transitions,
induce  interactions which are absent 
in conventional perturbation theory. 
In accord~\cite{th} with the general chiral anomaly relation,
instantons give rise to (hard) processes in which   
certain fermionic quantum numbers are violated, notably, 
chirality ($Q_{5}$) in (massless) QCD 
and baryon plus lepton number ($B+L$) in QFD. 

Implications of QCD-instantons for long-distance phenomena 
have been intensively studied in the past, 
mainly in the context of the phenomenological instanton 
liquid model~\cite{ss} and of lattice simulations~\cite{lattice}. Yet, a 
direct experimental verification of their existence is
lacking up to now. An experimental discovery of such a novel, 
non-perturbative manifestation of non-abelian gauge theories would 
clearly be of basic significance. 

The deep-inelastic regime is distinguished by the fact that here
hard QCD-in\-stan\-ton induced processes may both be {\it
calculated}~\cite{bb,mrs1,rs-pl} within in\-stan\-ton-per\-tur\-ba\-tion
theory  and possibly 
{\it detected experimentally}~\citer{rs,crs}.
As a  key feature it has recently been shown~\cite{mrs1}, that in DIS the 
generic hard scale ${\cal Q}$ cuts off instantons with {\it large size} 
$\rho\gg {\cal Q}^{-1}$, over which one has no control within 
perturbation theory in the instanton background.

\section{Cross-Sections}\label{s1}

\begin{figure}
\begin{center}
\epsfig{file=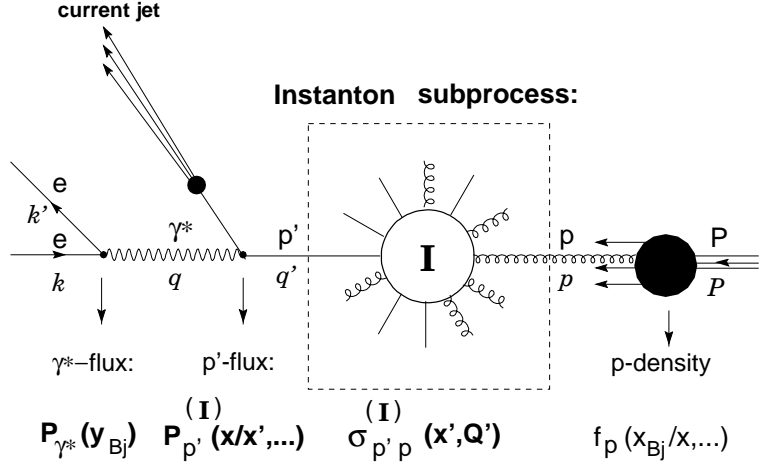,width=10cm}
\caption[dum]{\label{ev-displ}The leading instanton-induced process 
in the DIS regime of $e^\pm P$ scattering, violating chirality by
$\triangle Q_5=2\,n_f$.}
\end{center}
\end{figure}

The leading instanton ($I$)-induced process in the DIS regime of $e^\pm
P$ scattering is displayed in Fig.~\ref{ev-displ}. 
The dashed box emphasizes the  so-called  instanton-{\it subprocess}
with its own Bjorken variables,
        \begin{equation}
        Q^{\prime\,2}=-q^{\prime\,2}\geq 0;\hspace{0.5cm}
        \xpr=\frac{Q^{\prime\,2}}{2 p\cdot q^\prime}\le 1.
        \end{equation}         
The $I$-induced cross-section in unpolarized deep-inelastic 
$e^\pm P$ scattering, $d\sigma_{eP}$, can be shown to factorize~\cite{rs-pl} 
(in the Bjorken limit, $Q^2=-q^2$ large) into a sum of differential 
parton-parton luminosities, $d{\mathcal L}^{(I)}_{p^\prime p}$, and 
$I$-subprocess cross-sections, $\sigma_{p^\prime p}^{(I)}$, 
\begin{equation}
\label{ePcross}
\frac{d\sigma_{eP}^{(I)}}{d\xpr\,dQ^{\prime 2}} \simeq
\sum_{p^\prime,p}
\frac{d{\mathcal L}^{(I)}_{p^\prime p}}{d\xpr\,dQ^{\prime 2}}
\,\sigma_{p^\prime p}^{(I)}(\xpr,Q^{\prime 2}) .
\end{equation} 
Here $p^\prime =q^\prime,\overline{q}^\prime$ denotes the virtual 
quarks entering the $I$-subprocess from the photon side and
$p=q,\overline{q},g$ denotes the target partons. 
The differential luminosity $d{\mathcal L}^{(I)}_{p^\prime p}$, accounting
for the number of $p^\prime p$ collisions per $eP$ collision, has a 
convolution-like structure~\cite{rs}, involving integrations over the 
target-parton density, $f_p$, the $\gamma^\ast$-flux,
$P_{\gamma^\ast}$, and the {\it known}~\cite{dis97-phen} 
flux $P^{(I)}_{p^\prime}$ of the parton $p^\prime$ in the 
$I$-background. 

In Eq.~(\ref{ePcross}), the $I$-subprocess
total cross-section $\sigma_{p^\prime p}^{(I)}$ 
contains the essential instanton dynamics. We have 
evaluated the latter~\cite{rs-pl} 
by means of the optical theorem and the so-called  
$\iai$-valley approximation~\cite{valley} for the relevant 
$q^\prime g\Rightarrow q^\prime g$ forward elastic scattering
amplitude in the $\iai$ background. This method 
resums the exponentiating final state gluons in form of the known
valley action $S^{(\iai)}$ and reproduces standard $I$-perturbation theory at
larger  $\iai$ separation $\sqrt{R^2}$. While a semi-quantitative
evaluation of $\sigma_{p^\prime p}^{(I)}$ may also be performed by
means of standard $I$-perturbation theory, the use of the $\iai$-valley method 
allows to extend our calculations to a somewhat larger range in $\xpr$.

Corresponding to the symmetries of the theory, the instanton calculus
introduces at the classical level certain (undetermined) 
``collective coordinates'' like the $I\,(\overline{I})$-size parameters 
$\rho\,(\overline{\rho})$ and the $\iai$ distance 
$\sqrt{R^2/\rho\overline{\rho}}$ (in units of the size).  
Observables like $\sigma_{p^\prime p}^{(I)}$
must be independent thereof and thus involve integrations over all collective
coordinates. Hence, we have generically,
\begin{eqnarray}
\label{sigma}
        \sigma_{p^\prime p}^{(I)}
        &=&\int\limits_0^\infty {\rm d}\rho\,D(\rho)
        \int\limits_0^\infty {\rm d}\overline{\rho}\, 
        D(\overline{\rho})
        \int {\rm d}^4 R\ \ldots        
        \\
\nonumber
        &&\times\,
        {\rm e}^{-({\rho + \overline{\rho}})Q^\prime}
        {\rm e}^{i(p+q^\prime )\cdot R}
        {\rm e}^{-\frac{4\pi}{\alpha_s}(S^{(\iai)}(\xi)-1)}. 
\end{eqnarray}   

The first important quantity of interest, entering Eq.\,(\ref{sigma}), is 
the $I$-density, $D(\rho)$ (tunnelling amplitude).
It has been worked out a long time ago~\cite{th,morretal} in the framework of 
$I$-perturbation theory: (renormalization scale $\mu_r$)
\begin{eqnarray}
   D(\rho)
   &=&
   d \left(\frac{2\pi}{\alpha_s(\mu_r)}\right)^6
   \exp{(-\frac{2\pi}{\alpha_s(\mu_r)})}\frac{(\rho\, \mu_r)^b}{\rho^{\,5}},
   \label{density}
\\ 
\label{b-morretal}
b&=&\beta_0+\frac{\alpha_s(\mu_r)}{4\pi}
      (\beta_1-12\beta_0) ,
\end{eqnarray}
in terms of the QCD 
$\beta$-function coefficients, $\beta_0=11-\frac{2}{3}{n_f},\ 
\beta_1=102-\frac{38}{3} {n_f}$. In this form it satisfies 
renormalization-group invariance at the two-loop level~\cite{morretal}.
Note that the large, positive power $b$ of $\rho$ in the 
$I$-density~(\ref{density}) would make the integrations over the 
$I(\ai)$-sizes in Eq.~(\ref{sigma}) infrared divergent without the  
crucial exponential cut-off~\cite{mrs1} 
$e^{-(\rho+\overline{\rho})\,Q^\prime}$ arising from the virtual quark
entering the $I$-subprocess from the photon side. 

The second important quantity of interest, entering Eq.\,(\ref{sigma}), is 
the $\iai$-interaction, $S^{(\iai)}-1$. In the valley approximation, the 
$\iai$-valley action, 
$S^{(\iai)}\equiv \frac{\alpha_s}{4\pi}\,S[A_\mu^{(\iai)}]$, 
is restricted by conformal
invariance to depend only on the ``conformal separation'', 
$\xi=R^2/\rho\overline{\rho}+\rho/\overline{\rho}+\overline{\rho}/\rho$,
and its functional form is explicitly known~\cite{valley}. It is important
to note that the interaction between $I$ and $\ai$ remains {\it attractive}
for all separations $\xi$; specifically, the $\iai$-valley action 
decreases monotonically from 1 at infinite conformal separation to 
0 at $\xi =2$, which corresponds to $R^2=0$ and $\rho =\overline{\rho}$. 

The collective coordinate integration in the cross-section~(\ref{sigma})
can be performed via {\it saddle-point techniques}. One finds 
$R_\mu^\ast = (\rho^\ast \sqrt{\xi^\ast -2},\vec{0})$ and 
$\rho^\ast=\overline{\rho}^\ast$, where the saddle-point solutions $\rho^\ast$
and $\xi^\ast$ behave qualitatively as
\begin{equation}
\label{qual-exp}
\rho^\ast \sim \frac{4\pi}{\alpha_s Q^\prime};\hspace{4ex}
\sqrt{\xi^\ast -2} = \frac{R^\ast}{\rho^\ast} 
\sim 2\, \sqrt{\frac{\xpr}{1 -\xpr}} .
\end{equation}
Thus, the virtuality $Q^\prime$ controls the $I(\ai)$-size:  
As one might have expected intuitively, highly virtual partons probe only
small instantons. 
The Bjorken-variable $\xpr$, on
the other hand, controls the conformal separation between $I$ and $\ai$:
for decreasing $\xpr$, the conformal separation decreases.

\begin{figure}[t]
\vspace{-9pt}
\begin{center}
\epsfig{file=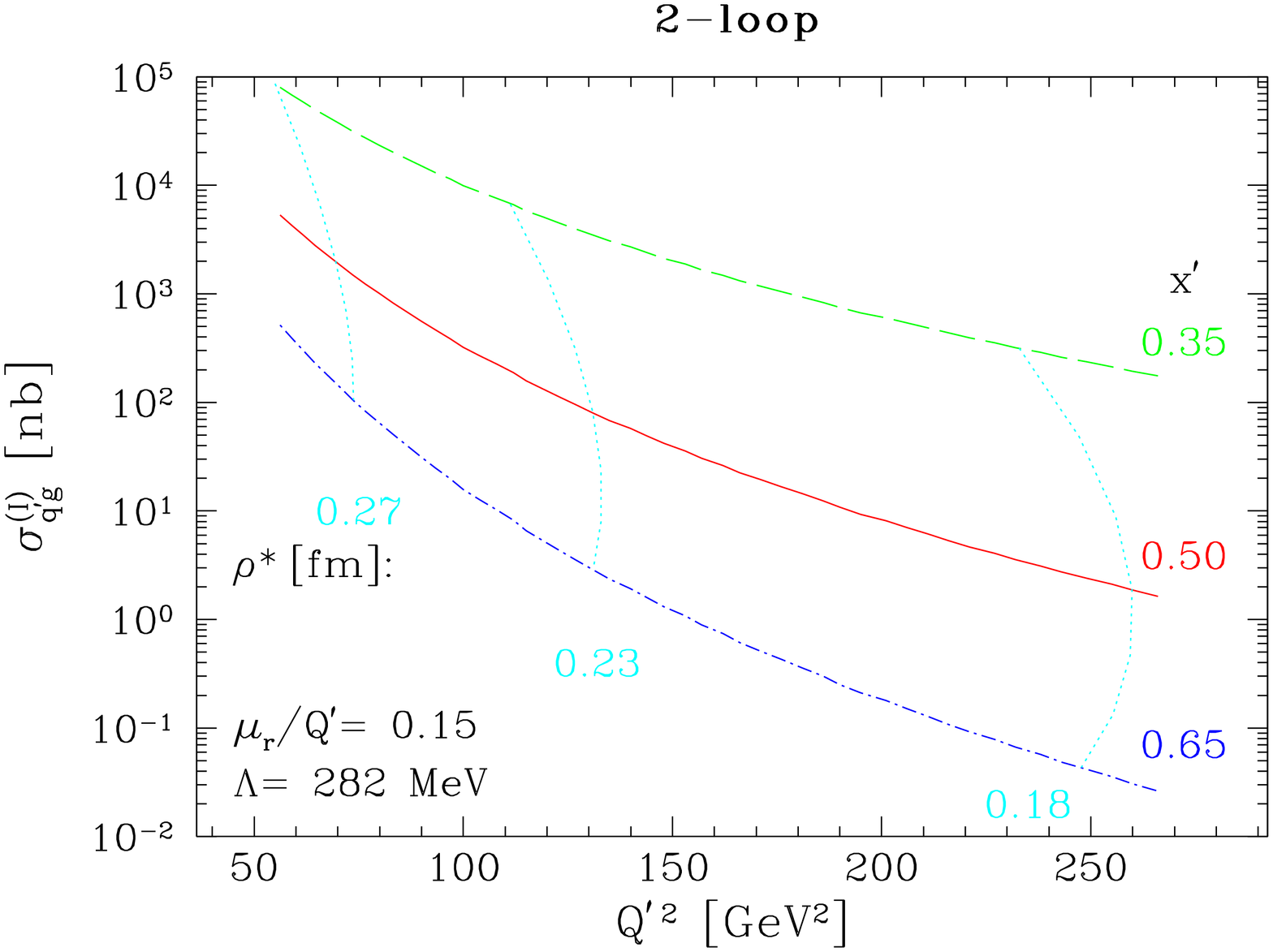,angle=-90,width=6.79cm}
\epsfig{file=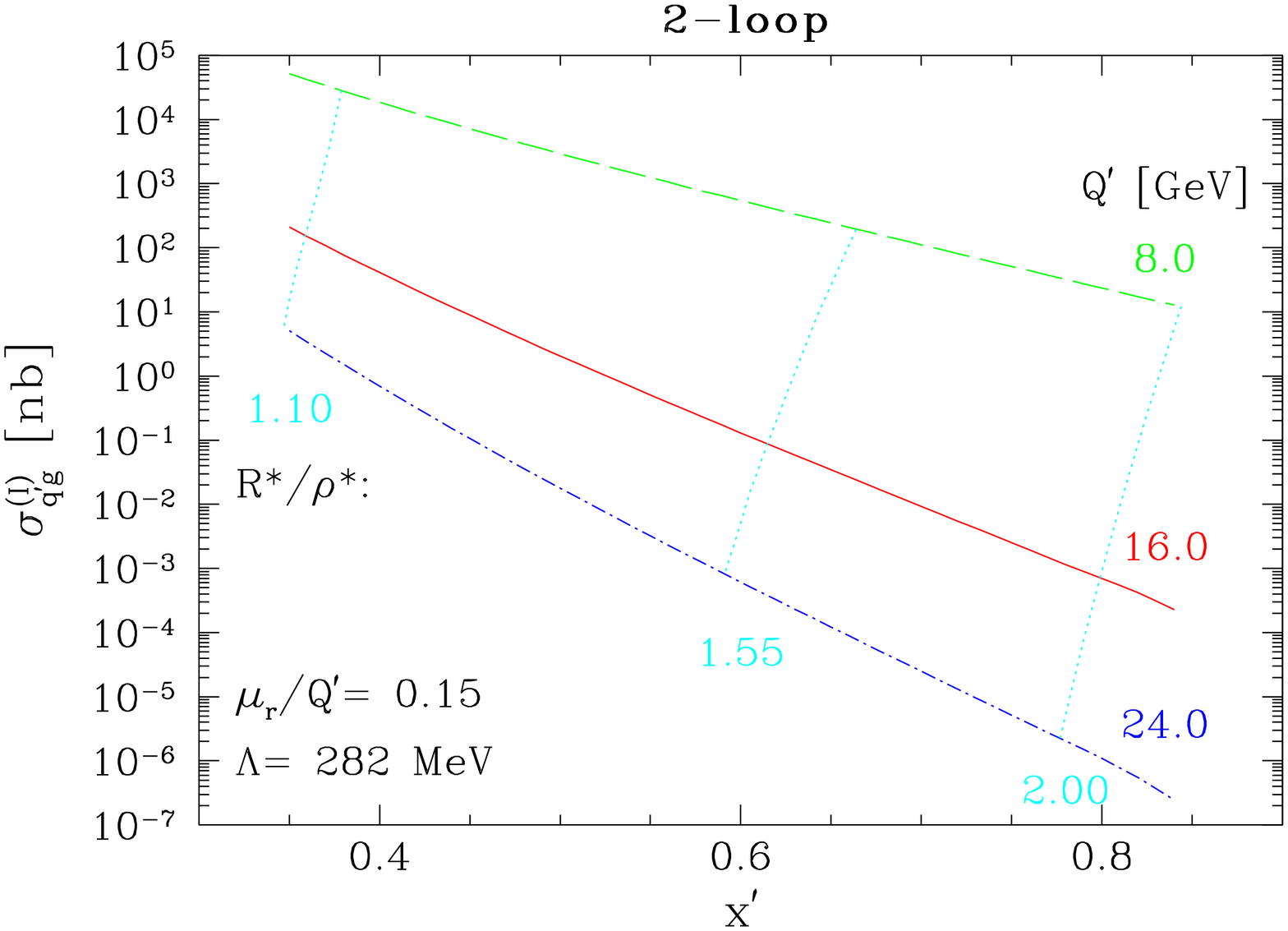,angle=-90,width=6.79cm}
\caption[dum]{\label{isorho}$I$-subprocess cross-section~\cite{rs-pl}.}
\end{center}
\end{figure}
Our quantitative results~\cite{rs-pl} on the {\it dominating} cross-section for
a target gluon, $\sigma^{(I)}_{q^\prime g}$, are shown in detail 
in Fig.~\ref{isorho}, both as functions of $Q^{\prime 2}$ (left) and of $\xpr$
(right). 
The dotted curves in Fig.~\ref{isorho}, indicating lines of constant 
$\rho^\ast$ (left) and of constant $R^\ast/\rho^\ast$ (right),    
nicely illustrate the qualitative relations~(\ref{qual-exp}) and their
consequences: the $Q^\prime$ dependence essentially maps the $I$-density, 
whereas the $\xpr$ dependence mainly maps the $\iai$-interaction.    

\begin{figure}[t]
\vspace{-16pt}
\begin{center}
\epsfig{file=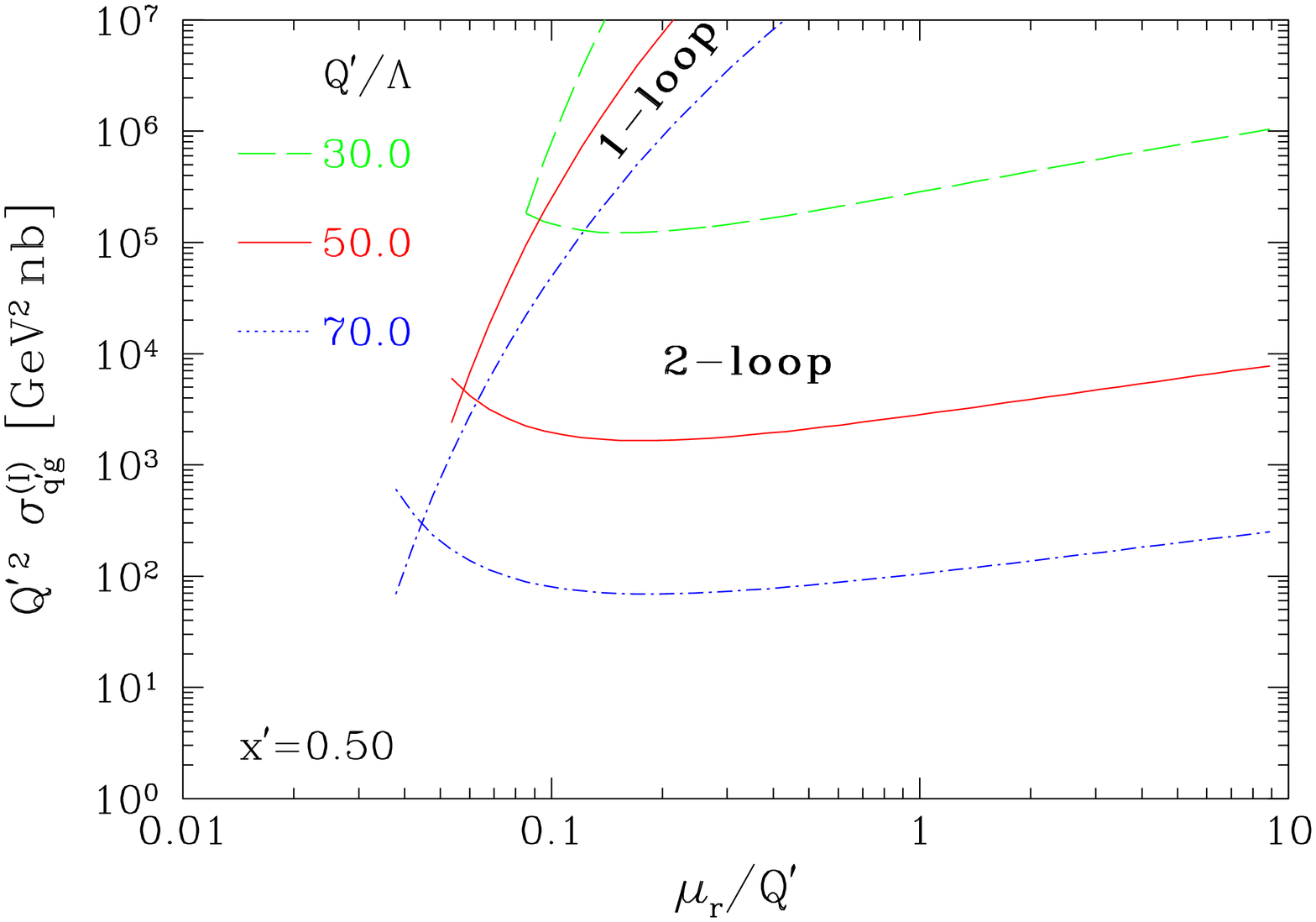,angle=-90,width=6.75cm}
\epsfig{file=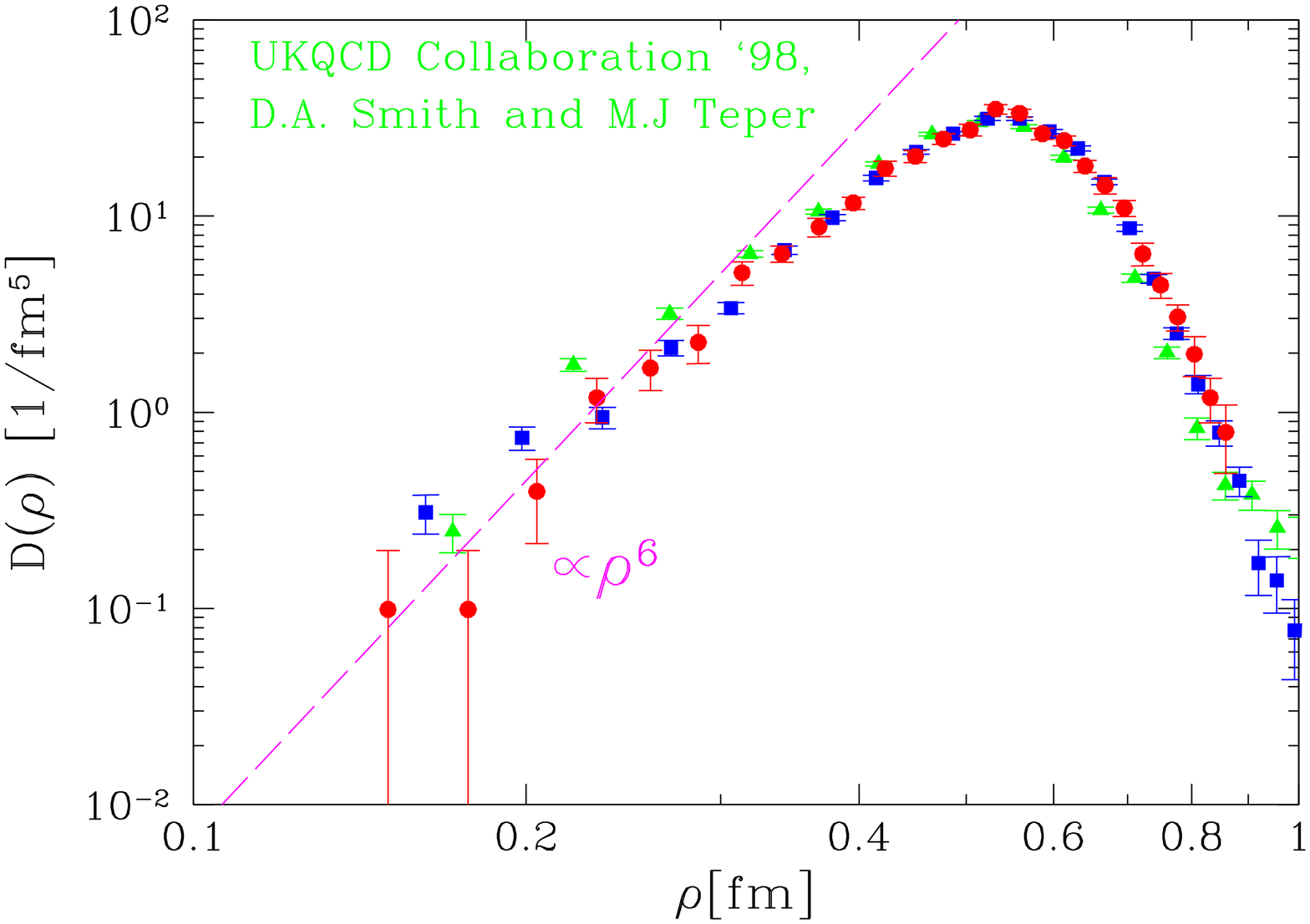,angle=-90,width=6.75cm}
\caption[dum]{\label{RGdep}{\it Left:} Renormalization-scale dependence of the 
$I$-subprocess cross-section for a target gluon~\cite{rs-pl}.  
{\it Right:} Support for the validity of $I$-perturbation theory for the
$I$-density $D(\rho)$ from recent lattice
data~\cite{rs-pl,ukqcd}.}
\end{center}
\end{figure}

Compared to our earlier estimates based on the one-loop
renormalization-group  invariant form~\cite{th} of the $I$-density $D(\rho)$,  
the residual dependence on the renormalization scale has now
dramatically weakened (c.f. Fig.~\ref{RGdep}~(left)). This important
stabilization of our predictions originates  from the use of the 
{\it two-loop} renormalization-group invariant improvement   
(\ref{density}, \ref{b-morretal}) of $D(\rho)$ from Ref.~\cite{morretal}.   

Intuitivelely one may expect~\cite{mrs1,bb} $\mu_r \sim 1/\langle \rho
\rangle \sim Q^\prime/\beta_0 ={\cal O}(0.1)\, Q^\prime$. Indeed, 
this guess turns out to match quite well
our actual choice of the  
 ``best'' scale, $\mu_r = 0.15\ Q^\prime$, determined by 
$\partial \sigma^{(I)}_{q^\prime g}/\partial \mu_r \simeq 0$ (c.f. 
Fig.~\ref{RGdep}~(left)).

Important information about the range of validity of 
$I$-perturbation for the $I$-density
and the $\iai$-interaction, in terms of the instanton collective coordinates 
($\rho\leq \rho_{\rm max}, R/\rho\geq (R/\rho)_{\rm min}$), can be obtained 
from recent (non-perturbative) lattice simulations of QCD and translated 
via the saddle-point relations~(\ref{qual-exp}) into a ``fiducial'' 
kinematical region 
($Q^\prime\geq Q^\prime_{\rm min},\xpr\geq x^\prime_{\rm min}$)~\cite{rs-pl}. 
In fact, from a comparison of the (one-loop) perturbative expression of the 
$I$-density~(\ref{density}) with recent lattice ``data''~\cite{ukqcd} 
one infers~\cite{rs-pl} semi-classical $I$-perturbation theory to be valid 
for $\rho \lwig \rho_{\rm max}\simeq 0.3$ fm (c.f. Fig.~\ref{RGdep}~(right)).
Similarly, it can be argued~\cite{rs-pl} that the attractive, semi-classical 
valley result for the $\iai$-interaction is supported by lattice
simulations  down to a minimum conformal separation $\xi_{\rm
min}\simeq 3$, i.e.  
$(R^\ast/\rho^\ast)_{\rm min}\simeq 1$. The corresponding 
``fiducial'' kinematical region for our 
cross-section predictions in DIS is then obtained as~\cite{rs-pl}
\begin{equation}
 \left.\begin{array}{lcl}\rho^\ast&\lwig&  
         0.3 {\rm\ fm};\\[1ex]
 \frac{R^\ast}{\rho^\ast}&\gwig&
  1\\
 \end{array}\right\}\Rightarrow
 \left\{\begin{array}{lclcl}Q^\prime&\geq &Q^\prime_{\rm min}&\simeq&
 8 {\rm\ GeV};\\[1ex]
 x^\prime&\geq &x^\prime_{\rm min}&\simeq &0.35.\\
 \end{array} \right .
\label{fiducial}
\end{equation}

\begin{figure}[t]
\vspace{-25pt}
\begin{center}
\epsfig{file=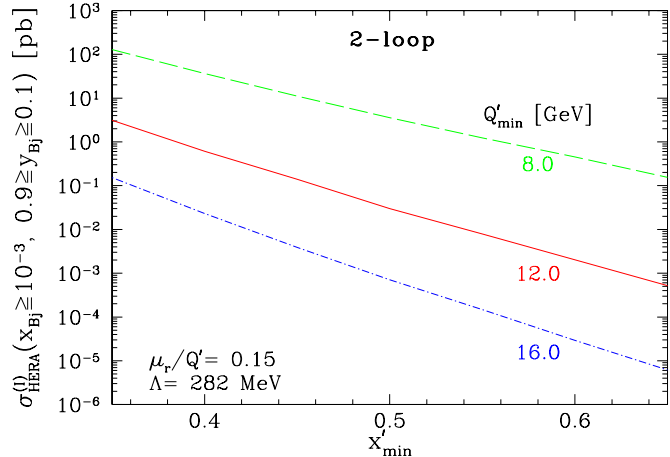,angle=-90,width=10cm}
\caption[dum]{\label{sigHERA}$I$-induced cross-section at HERA~\cite{rs-pl}.}
\end{center}
\end{figure}
Fig.~\ref{sigHERA} displays the finalized $I$-induced cross-section at HERA,
as function of the cuts $x^\prime_{\rm min}$ and
$Q^\prime_{\rm min}$, as
obtained with the new release ``QCDINS 1.6.0''~\cite{crs} of our $I$-event
generator. For the following ``standard cuts'', 
\begin{eqnarray}
\label{standard-cuts} 
{\mathcal C}_{\rm std}=
\xpr\ge0.35,Q^\prime\ge 8\, {\rm GeV}, x_{\rm Bj}\geq 10^{-3},
0.1\leq y_{\rm Bj}\leq 0.9 ,
\end{eqnarray}
including the minimal cuts~(\ref{fiducial}) extracted from lattice 
simulations, we specifically obtain 
\begin{eqnarray}
\label{minimal-cuts}
\sigma^{(I)}_{\rm HERA}({\mathcal C}_{\rm std}) 
&=& 
126^{\,+300}_{\,-100}\ {\rm pb},
\end{eqnarray}
where the uncertainties result mainly from the experimental uncertainty
in the QCD scale $\Lambda$.
Hence, with the total luminosity accumulated by experiments at HERA, 
${\mathcal L}={\mathcal O}(80)$ pb$^{-1}$, there should be already 
${\mathcal O}(10^4)$ $I$-induced events from the kinematical 
region~(\ref{standard-cuts})
on tape. Note also that the cross-section quoted in Eq.~(\ref{minimal-cuts})
corresponds to a fraction of $I$-induced to normal DIS (nDIS) events of
\begin{equation}
\label{fraction-fiducial}
f^{(I)}({\mathcal C}_{\rm std}) = 
\frac{\sigma_{\rm HERA}^{(I)}({\mathcal C}_{\rm std})}
{\sigma_{\rm HERA}^{({\rm nDIS})}({\mathcal C}_{\rm std})}
=
{\mathcal O}(1)\, \%.
\end{equation} 

Thus, it appears to be a question of {\it signature} rather than a 
question of {\it rate} to discover $I$-induced scattering processes at HERA. 
Hence, we turn next to the {\it final states} of $I$-induced events in DIS.

\section{Searches at HERA}\label{s2}

A Monte-Carlo generator, QCDINS,  for instanton-induced events in DIS, 
interfaced to HERWIG, has been developed \citer{grs,crs}. 
The Monte-Carlo simulation essentially proceeds in
three steps. First, quasi-free partons are produced by QCDINS with the
distributions prescribed by the hard process matrix elements. Next, 
these primary partons give rise to parton showers, as described
by HERWIG. Finally, the showers are converted into hadrons, again
within HERWIG. 

\begin{figure}[t]
\begin{center}
\epsfig{file=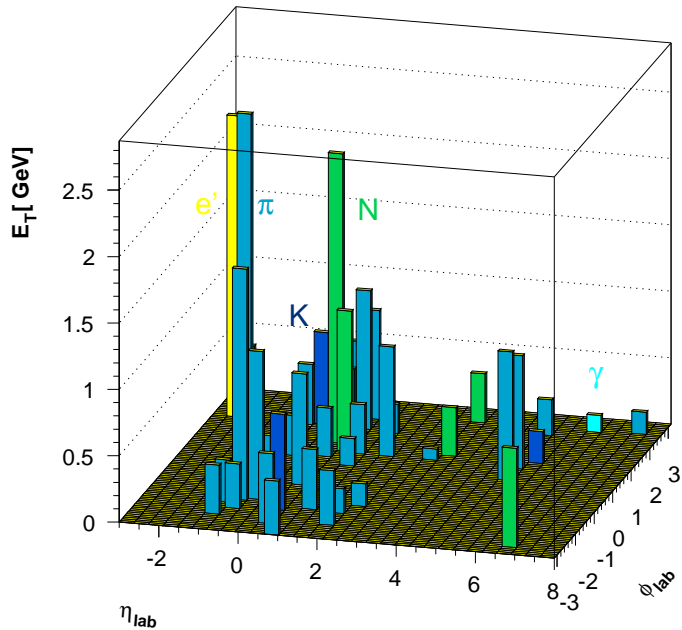,width=9cm}
\caption[dum]{\label{typical}Lego plot of a typical $I$-induced event
in the HERA lab system.}
\end{center}
\end{figure}
In Fig.~\ref{typical} we display the lego plot of a typical $I$-induced event 
at HERA, as generated by QCDINS. Its characteristics directly reflect 
the essential features of the underlying $I$-subprocess and thus may
be intuitively understood: 

After hadronization, the current quark in Fig.~\ref{ev-displ} gives
rise to a current-quark jet. No further distinct qjets are expected, since the
partons  from  the $I$-subprocess are emitted
spherically  symmetric in the   
$p^{\prime}p$ c.m. system (``$I$-c.m. system''). The gluon multiplicities are 
generated according to a Poisson distribution with mean multiplicity
$\langle n_g \rangle^{{(I)}}\sim 1/\alpha_s\sim 3$. In view of the large
required chirality violation $\Delta Q_5=2 n_f=6$,
the total mean parton multiplicity is large, of the order of ten. 
After hadronization, we therefore expect from the $I$-subprocess a
final state structure reminiscent of a decaying fireball: 
$\gwig 20$ hadrons are produced, always including {\it strange} ones. 
They are concentrated in 
a ``band'' at fixed pseudorapidity $\eta$ in
the ($\eta$, azimuth angle $\phi$)-plane. Due to the boost 
from the $I$-c.m. system to the HERA-lab system, the center of the
band is shifted away from $\eta=0$. Its width is 
of order $\Delta\eta \simeq 1.8$, as typical for a spherically
symmetric  event. 
For $\xpr\simeq 0.35$ and $Q^{\prime}\simeq 8$ GeV, the total
invariant mass  of the $I$-system,
$\sqrt{s^{\prime}}=Q^{\prime}\sqrt{1/\xpr -1}$, 
is expected to be in the 10 GeV range. All these expectations are clearly
reproduced by our Monte-Carlo simulation.

These features have been exploited by experimentalists at HERA to place 
first upper limits on the fraction of $I$-induced events to normal DIS 
events, in a similar kinematical region as our standard 
cuts~(\ref{standard-cuts}): From the search of a $K^0$ excess in the ``band'' 
region, the H1 Collaboration could establish a limit of 
$f_{\rm lim}^{(I)}=6$ \%, while the search of an excess in charged 
multiplicity yields $f_{\rm lim}^{(I)}=2.7$ \%~\cite{h1-limits}.
The limit from the charged multiplicity distribution has been further 
improved in Ref.~\cite{ck-limits} to about 1 \%.   

Since these experimental upper limits already range close to our 
estimate~(\ref{fraction-fiducial}), it becomes extremely interesting to 
investigate further possibilities to discriminate $I$-induced from nDIS 
final states. A dedicated multi-observable analysis is in progress,
with the aim of producing an instanton-enriched data sample.
Furthermore, strategies to reconstruct $\xpr$ and $Q^{\prime 2}$ from 
the final state are being developped.   

Altogether, a {\it decisive} search for $I$-induced at HERA
appears to be feasible.

\end{document}